\documentclass[12pt]{iopart}

\usepackage{cite}
\usepackage{graphicx}
\usepackage{subfig}
\usepackage{epstopdf}
\usepackage{color}
\usepackage{lineno}
\begin{document}

\title{Microwave spectroscopic study of the hyperfine structure of antiprotonic $^3$He}

\author{
S Friedreich$^{1}$, 
D~Barna$^{2,3}$,
F~Caspers$^{4}$,
A~Dax$^{2}$,
R~S~Hayano$^{2}$,
M~Hori$^{2,5}$,
D~Horv\'ath$^{3,6}$,
B~Juh\'asz$^{1}$\footnote{Present address: Lufthansa Systems Hung{\'a}ria Kft., Neumann J{\'a}nos utca 1/E, H-1117 Budapest},
T~Kobayashi$^{2}$,
O~Massiczek$^{1}$,
A~S\'ot\'er$^{5}$,
K~Todoroki$^{2}$,
E~Widmann$^{1}$ and
J~Zmeskal$^{1}$
}

\address{$^{1}$Stefan Meyer Institute for Subatomic Physics, Austrian Academy of Sciences, Boltzmanngasse 3, A-1090 Vienna, Austria}
\address{$^{2}$Department of Physics, University of Tokyo, 7-3-1 Hongo, Bunkyo-ku, Tokyo 113-0033, Japan}
\address{$^{3}$Wigner Research Centre for Physics, Institute for Particle and Nuclear Physics, H-1121 Budapest, Konkoly-Thege 29-33, Hungary}
\address{$^{4}$CERN, CH-1211 Geneva, Switzerland}
\address{$^{5}$Max-Planck-Institut f\"{u}r Quantenoptik, Hans-Kopfermann-Strasse 1, D-85748 Garching, Germany}
\address{$^{6}$Institute of Nuclear Research of the Hungarian Academy of Sciences, H-4001 Debrecen, PO Box 51, Hungary}
\ead{susanne.friedreich@oeaw.ac.at}

\begin{abstract}
In this work we describe the latest results for the measurements of the hyperfine structure of antiprotonic $^3$He. 
Two out of four measurable super-super-hyperfine {SSHF} transition lines of the $(n,L)=(36,34)$ state of antiprotonic $^3$He were observed. The measured frequencies of the individual transitions are $11.12548(08)$~GHz and $11.15793(13)$~GHz, with an increased precision of about 43\% and 25\% respectively compared to our first measurements with antiprotonic $^3$He 
[S. Friedreich {\it et al.}, Phys. Lett. B 700 (2011) 1--6].
They are less than 0.5~MHz higher with respect to the most recent theoretical values, still within their estimated errors. Although the experimental uncertainty for the difference of $0.03245(15)$~GHz between these frequencies is large as compared to that of theory, its measured value also agrees with theoretical calculations. The rates for collisions between antiprotonic helium and helium atoms have been assessed through comparison with simulations, resulting in an elastic collision rate of $\gamma_e = 3.41 \pm 0.62$~MHz and an inelastic collision rate of $\gamma_i = 0.51 \pm 0.07$~MHz. 
\end{abstract}

\begin{center} {\footnotesize Version \today}  \end{center}

\pacs{36.10.-k, 32.10.Fn, 33.40.+f}
\submitto{Journal of Physics B}  

\section{Introduction}
\begin{linenumbers}
\textit{Antiprotonic helium} $\overline{\mathrm{p}}$He$^{+}$ is a metastable three-body system consisting of one electron in the ground state, the helium nucleus and one antiproton~\cite{Iwasaki,Yamazaki:93,Yamazaki:02,Hayano:2007}. This \textit{exotic} atom can be created whenever an antiproton in the vicinity of a helium atom is slowed down to its ionization energy of $\sim$24.6~eV or below. The antiproton can eject one of the two electrons from the ground state and replace it. The antiproton is captured and, due to its high mass, most likely to be in states with high angular momentum and with principal quantum number $n = n_0 \equiv \sqrt{M^{*}/m_{\mathrm{e}}} \sim 38$ , $M^{*}$ being the reduced mass of the system, while the electron remains in the ground state. Therefore, these newly-formed atoms occupy circular states with $L$ close to $n$, where $L$ is the angular momentum quantum number.  

A majority of 97\% of these exotic atoms find themselves in states dominated by Auger decay. Due to the Auger excitation of the electron they ionize within a few nanoseconds after formation. 
The remaining 3\% of antiprotonic helium atoms remain in metastable, radiative decay-dominated states. 
These states are relatively long lived, having a lifetime of about 1-2~$\mu$s -- a time window that can be used to do laser and microwave spectroscopy measurements~\cite{Pask:2009,friedreich:2011,nature}. 
\end{linenumbers}

\section{Hyperfine structure of antiprotonic helium}
\begin{figure}
\centering
\includegraphics[width=0.47\textwidth, trim=0 0 0 0,clip]{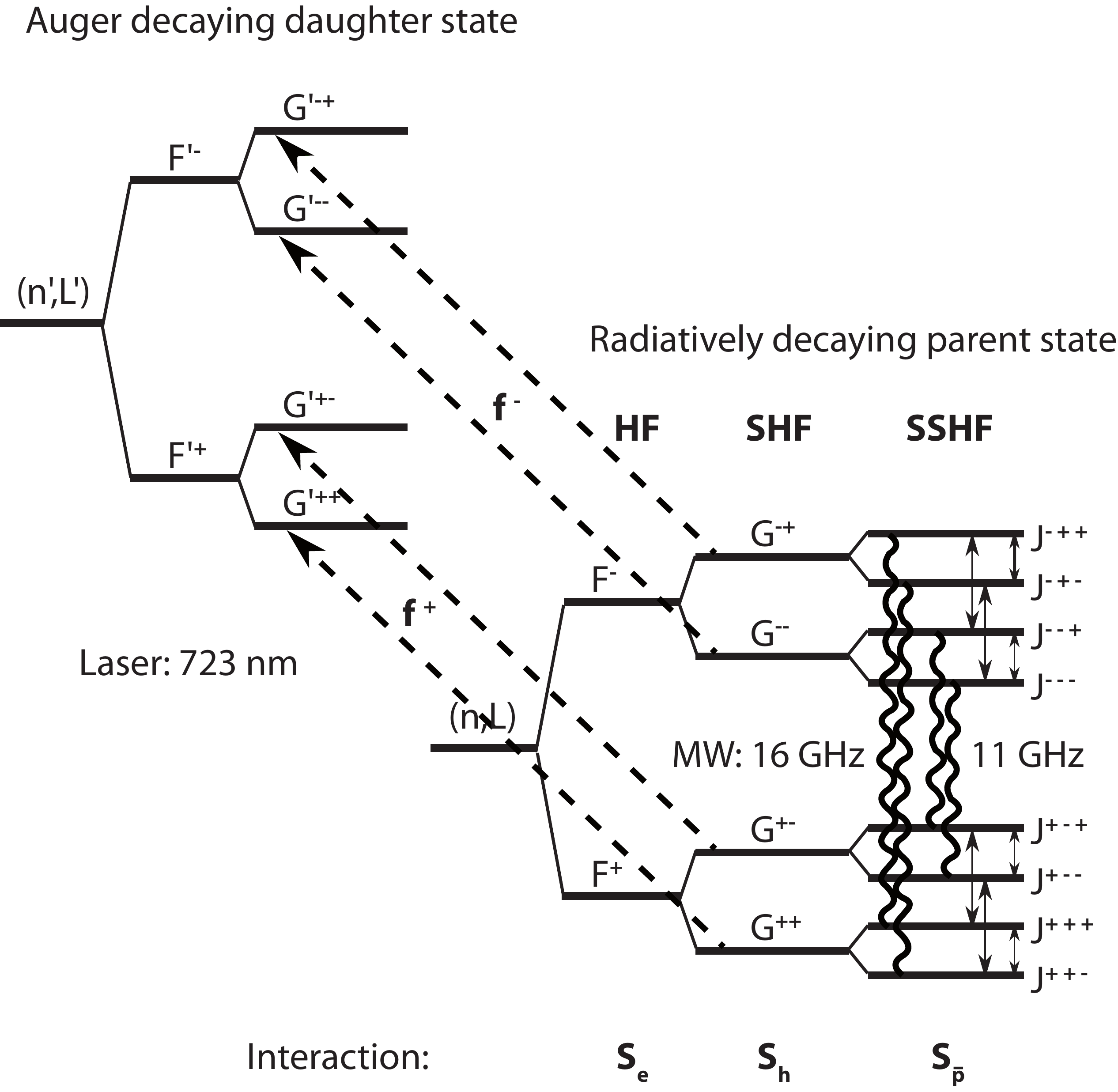}
\caption{\small{A schematic drawing of the laser-microwave-laser method. The dashed arrows indicate the laser transitions between the {SHF} levels of the radiative decay-dominated state $(n,L)=(36,34)$ and the Auger decay-dominated state $(37,33)$ of $\bar{\mathrm{p}}^{3}$He$^{+}$. The wavy lines illustrate the microwave-induced transitions between the {SSHF} levels of the long-lived state.}} \label{fig:He3_lMWl}
\end{figure}

\begin{linenumbers}
The interaction of magnetic moments between electron, antiproton and helium nucleus gives rise to a splitting of the $\overline{\mathrm{p}}^{3}$He$^{+}$ energy levels. The coupling of the electron spin $\vec{S}_{\mathrm{e}}$ and the orbital angular momentum of the antiproton $\vec{L}$ leads to the primary splitting of the state into a doublet structure, referred to as \textit{hyperfine ({HF}) splitting}. The angular momentum $\vec{F}=\vec{L}+\vec{S}_\mathrm{e}$ defines the two substates with quantum numbers $F_+=L+\frac{1}{2}$ and $F_-=L-\frac{1}{2}$. The non-zero spin of the $^{3}$He nucleus causes a further, so-called \textit{super-hyperfine ({SHF}) splitting}, which can be characterized by the angular momentum $\vec{G}=\vec{F}+\vec{S}_{\mathrm{h}}=\vec{L}+\vec{S}_{\mathrm{e}}+\vec{S}_\mathrm{h}$, where $\vec{S}_\mathrm{h}$ is the spin of the helium nucleus. This results in four {SHF} substates. At last, the spin-orbit interaction of the antiproton orbital angular momentum and antiproton spin $\vec{S}_{\bar{\mathrm{p}}}$ in combination with the contact spin-spin and the tensor spin-spin interactions between the particles result in a further splitting of the {SHF} substates into eight substates which we call \textit{super-super-hyperfine ({SSHF}) splitting}. This octuplet structure can be described by the angular momentum $\vec{J}=\vec{G}+\vec{S}_{\bar{\mathrm{p}}}=\vec{L}+\vec{S}_{\mathrm{e}}+\vec{S}_{\mathrm{h}}+\vec{S}_{\bar{\mathrm{p}}}$. Even though the magnetic moment of the antiproton is larger than that of the $^{3}$He nucleus, the former has a smaller overlap with the electron cloud. Therefore it creates a smaller splitting. The complete hyperfine structure for $\overline{\mathrm{p}}^{3}$He$^{+}$ is illustrated in Fig.~\ref{fig:He3_lMWl}. 
\linebreak

The interest in $\overline{\mathrm{p}}^3$He$^+$ arises from an additional contribution to the hyperfine structure caused by the coupling of the nuclear spin to the antiproton orbital momentum with respect to	 $\overline{\mathrm{p}}^4$He$^+$~\cite{Pask:2009}. Such a measurement would allow a more rigorous test of {QED} theory. The accurate knowledge of the hyperfine structure of antiprotonic helium is essential for the calculation of the laser transition energies at the level of ppb accuracy needed for comparison to laser spectroscopy experiments and the extraction of the antiproton-to-electron mass ratio \cite{nature}. An experimental verification of the HFS splitting in $\overline{\mathrm{p}}^3$He$^+$ is therefore of great importance.

The calculations of the hyperfine structure were developed by two different groups~\cite{Bakalov:98,Korobov:01,Yamanaka:01,Kino:03APAC,Korobov:06}. 
This series of experiments, studying the $(n,L)=(36,34)$ state, was the first attempt to measure the microwave transition frequencies between hyperfine substates of $\overline{\mathrm{p}}^3$He$^+$. Transitions between the {SSHF} states were induced by a magnetic field oscillating in the microwave frequency range. Due to technical limitations of the microwave input power, only the transitions which flip the spin of the electron could be measured. There are four such "allowed" {SSHF} transitions for the $(n,L)=(36,34)$ state of $\overline{\mathrm{p}}^3$He$^+$ two of which we investigated with the present work:
\begin{eqnarray}     
\nu_{\mathrm{HF}}^{--}~:~J^{---}=L-\frac{3}{2}~\longrightarrow ~J^{+--}=L-\frac{1}{2} \nonumber   \\
\nu_{\mathrm{HF}}^{-+}~:~J^{--+}=L-\frac{1}{2}~\longrightarrow ~J^{+-+}=L+\frac{1}{2}  \nonumber   
\label{MWtrans_He3}         
\end{eqnarray}

\section{Laser-microwave-laser spectroscopy}
\label{sec:method}
\begin{linenumbers}
The first observation of a hyperfine structure in antiprotonic helium was achieved in a laser scan of the $(n,L)=(37,34)\rightarrow(38,35)$ transitions in  $\overline{\mathrm{p}}^4$He$^+$ \cite{Widmann:1997}. Due to the limited precision achievable in a laser scan, a \textit{laser-microwave-laser method} (Fig.~\ref{fig:He3_lMWl}) was introduced in \cite{Widmann:02}. It is based on a three-step process involving laser and microwave stimulated resonance transitions.

After antiprotonic helium is formed, the atoms in the hyperfine substates are all equally populated. Therefore at first a population asymmetry between the {SSHF} substates of the measured radiative decay state $(n,L)$ needs to be created. This depopulation is induced by a short laser pulse which transfers the majority of antiprotons from one of the {HF} states of the radiative decay-dominated, metastable parent state to an Auger decay-dominated, short-lived daughter state. In this experiment the $f^+$ transition is used. The bandwidth of the laser (100~MHz) and Doppler broadening at 6 K (300~MHz) are small enough compared to the difference of $f^{-}-f^{+}\sim 1.7$~GHz  so that the $f^-$ transition is not affected and a population asymmetry can be achieved. 
The antiprotons in the short-lived state annihilate within a few nanoseconds. In the next step, a microwave frequency pulse tuned around the transition frequency between two {SSHF} substates of the metastable state is applied. If the microwave field is on resonance with one of these transitions, it will cause a population transfer and thus partial refilling of one of the previously depopulated states. A second laser pulse will then again cause depopulation of the same {HF} substate and subsequently Auger decay of the transferred atoms and annihilation of the antiprotons in the nucleus will occur. The number of annihilations after the second laser pulse will be the larger the more antiprotons were transferred by the microwave pulse. 
\linebreak

\begin{figure}
\centering
\includegraphics[width=0.47\textwidth, trim=0 0 0 0,clip]{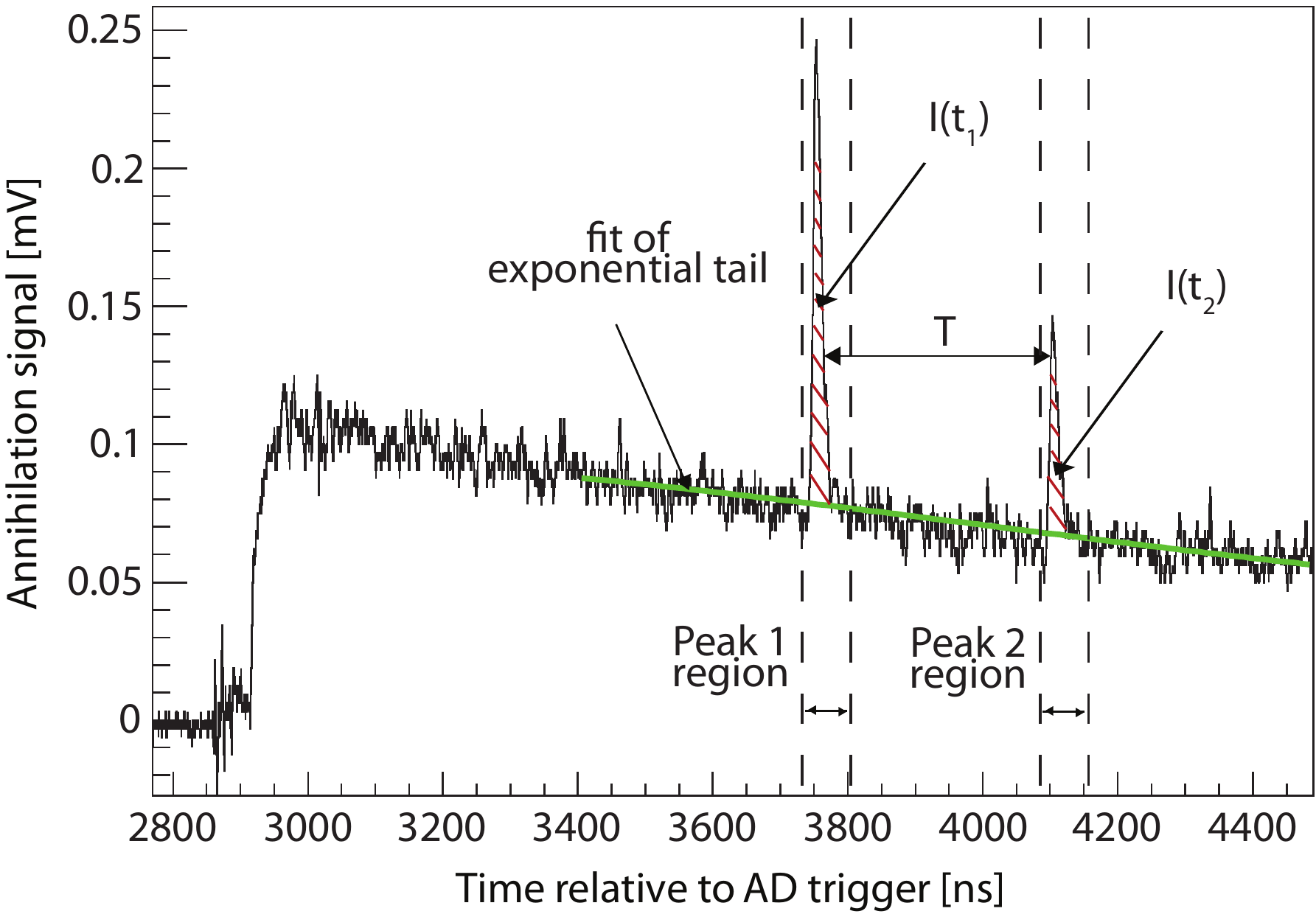} 
\caption{\small{A part of the analog delayed annihilation time spectrum ({ADATS}) with the two laser-stimulated annihilation peaks against the exponential decaying background of the metastable cascade.  $T$ denotes the delay time between the two laser pulses. The photomultipliers of the Cherenkov counters used to record this spectrum are gated off during the initial $\overline{\mathrm{p}}$ pulse arrival~\cite{Hori:2003}. Thus, the prompt peak is cut off below 2900 ns and only the annihilations due to the metastable state depopulation are recorded.}} \label{fig:Adats}
\end{figure}

When the antiprotons first enter the helium gas, a large annihilation peak (''prompt peak'') is caused by the majority of formed $\overline{\mathrm{p}}\mathrm{He}^{+}$ atoms which find themselves in Auger decay-dominated states and annihilate within picoseconds after formation. At later times, this peak exhibits an exponential tail due to $\overline{\mathrm{p}}\mathrm{He}^{+}$ atoms in the metastable states cascading more slowly towards the nucleus. This constitutes the background for the laser-induced annihilation signals. 
The daughter state has a very short lifetime of $\sim$10~ns and thus the population transfer is indicated by a sharp annihilation peak against the background during the two laser pulses. The area under these peaks is proportional to the population transferred to this short-lived state. This spectrum, with the two laser-induced peaks super-imposed on the exponential tail -- as displayed in Fig.~\ref{fig:Adats} -- is called \textit{analogue delayed annihilation time spectrum} or {ADATS}.
\linebreak

Since the intensity of the antiproton pulse fluctuates from shot to shot, the peaks must be normalised by the total intensity of the pulse (total). This ratio is referred to as \textit{peak-to-total}. The peak-to-total (ptt) corresponds to the ratio of the peak area ($I(t_1)$ or $I(t_2)$) to the total area under the full spectrum. If the second laser annihilation peak is further normalised to the first one, the total cancels out. The frequencies of the two {SSHF} transitions can now be obtained as distinct lines by plotting $I(t_2)/I(t_1)$ as a function of the microwave frequency. The ratio $I(t_2)/I(t_1)$ is largely independent of intensity and position fluctuations of the antiproton beam. The intensity of the transition lines is subject to the time delay between the two laser pulses and thus also to collisional relaxation processes~\cite{Kman, Korenman:2009, Korenman:2012, Friedreich:2012}. This means that, once the first laser has caused depopulation, the system will start to relax through spin exchanging collisions between antiprotonic helium atoms and regular helium atoms. Refilling from higher-lying states also contributes to the equalization of the hyperfine substate populations. In general, a short delay $T$ is preferable because the signal height will decrease for longer laser delay times as a result of the exponential decay of the metastable state populations. However, the linewidth of the {RF} transition will increase if the delay is too short. 
Further, far higher {RF} power will be required to complete one spin-flip. If the delay is too long, the collisional relaxation of the system would already have eliminated any asymmetry between the two states caused by the first laser pulse. The signal would be too low to be observed.

The two pulsed lasers were fixed to a wavelength of 723.877~nm, with a pulse length of 8-12~ns, to induce the $f^{+}$ laser transition between the $(n,L)=(36,34)$ and the $(n',L')=(37,33)$ state. The pulse length should be comparable to or longer than the Auger lifetime of the short-lived state. Generally spoken, the longer the laser pulse the larger the achieved depopulation and thus the resulting annihilation signal. 
The depopulation also depends on the laser pulse energy. It is important to find the appropriate laser fluence where the power is saturated and therefore the laser depletion efficiency is optimized in order to avoid power broadened resonance lines and as a consequence partial depopulation of the other {HF} transition line f$^{-}$.
For this experiment a pulse-amplified continuous-wave laser system with a narrow linewidth of about 100~MHz was used~\cite{Hori:2006}. The laser fluence was in the range of 20-40~mJ$/$cm$^{2}$, the laser waist $\sim$5~mm, leading to a depletion efficieny of about 90\% -- based on numerical simulations of the laser transition processes~\cite{Friedreich:2012}. 

There are several limitations to the choice of the measured state, such as availability of a laser source in the required frequency range or the splitting of the transitions between the {HF} states of the daughter and the parent state. The laser transition between the $(n,L)=(36,34)$ and the $(n',L')=(37,33)$ state was chosen because it is easily stimulated and the primary population is large, thus leading to a large signal. The captured fraction of antiprotons in the measured metastable state $(n,L)=(36,34)$ is (3-4)$\times 10^{-3}$~\cite{Hori:02}.
\end{linenumbers}

\section{Experimental setup}
\label{sec:expset}

\begin{figure*}[b]
\centering
\includegraphics[width=0.35\textwidth, trim=0 0 0 0,clip]{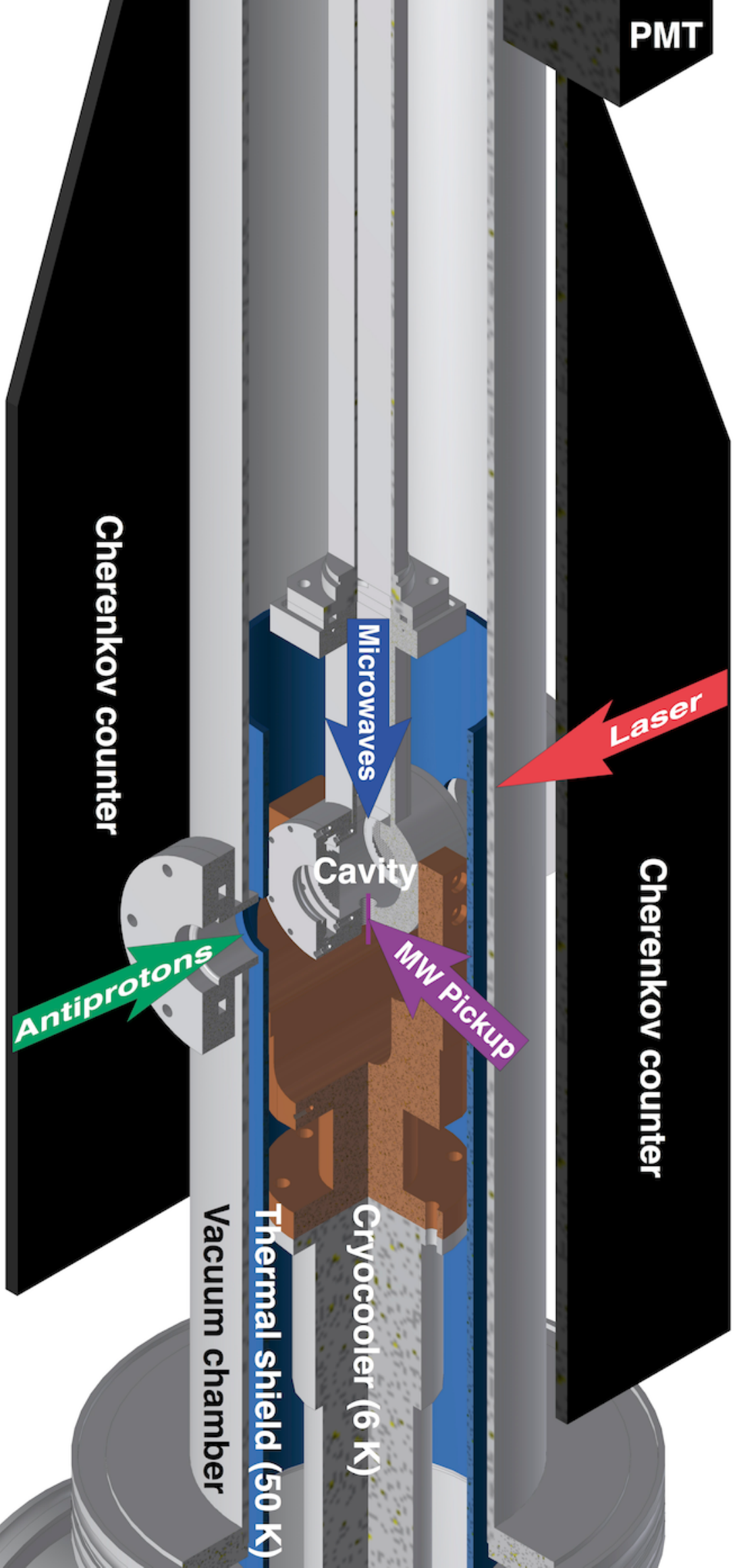}
\caption{\small{Central part of the experimental setup. Antiproton and laser beams coming from opposite sides are injected into the microwave cavity which also contains the helium gas. Microwaves are fed through a wave guide from top, and the microwave power is measured by a small antenna. Outside the vacuum chamber two Cherenkov counters are mounted to detect the pions resulting from the annihilations.}}
\label{fig:NewCryostat}
\end{figure*}

The antiprotons for the experiment are provided by the {AD} (Antiproton Decelerator) at {CERN}~\cite{Maury:1997}, with a pulsed beam of (1-3)$\times 10^7$ antiprotons at an energy of 5.3~MeV, a pulse length of 100-300 ns, and a repetition interval of about 100~s. The particles are stopped in a helium gas target, with a gas pressure of 250~mbar, cooled down to a temperature of about 6~K. This target is built as a cylindrical chamber whose axis is parallel to the beam direction and which is designed to act also as a microwave cavity resonating in the TM$_{110}$ mode. The faces of the cylindrical cavity have a 25 $\mu$m thick titanium window for the antiproton beam and a 4 mm thick fused silica window for the laser beam to enter \cite{massiczek:2011}, and are equipped with meshes to contain the microwaves. 

In order to measure the annihilation decay products two Cherenkov counters are mounted around the target volume, connected to photomultipliers (cf. Fig.~\ref{fig:NewCryostat}) . They are gated off during the initial $\overline{\mathrm{p}}$ pulse arrival~\cite{Hori:2003} in order to count only the photons arriving from the induced annihilations. A vector network analyzer ({VNA}, Rhode \& Schwarz {ZVB}20) synthesizes the microwave pulse that is further amplified by a travelling wave tube amplifier ({TWTA}, {TMD} {PTC}6358) from where a waveguide system transmits the pulse of $\sim$20~$\mu$s length to the cavity. The waveguide is over-coupled to the cavity resulting in a low quality factor of $Q=160$. The frequency of the microwave radiation is tuned by changing the frequency of the VNA, increasing the input power off-resonance so to keep the power inside the cavity constant. 
The microwave power inside the cavity is measured by a pickup antenna and a calibrated diode (Agilent 8474B). Input powers of maximum 40 W were used to achieve a constant power of 7.5 W inside the cavity. A detailed discussion on the microwave apparatus, including design, simulation, construction and calibration, can be found in~\cite{friedreichNIM:2012}.

A cryostat with compressor-based cooling system was built to cool the experimental apparatus without abundant use of coolants, to allow an efficient cooling procedure and thus little loss of measurement time. The microwave cavity is filled with helium gas and cooled down directly to about 6~K by mounting it on a coldhead~\cite{massiczek:2011}.  By use of additional degrader foils (Polymide film foils~\footnote{Upilex foil made by {UBE Industries}} of about 70~$\mu$m thickness) the antiprotons could be stopped in the center of a cavity in a volume of about 1~cm$^{2}$ \cite{Sakaguchi:04}.

\section{Results}
\label{sec:res}
In preparation for the actual investigation of the hyperfine substructure, via microwave resonance, several studies are required to optimize the parameters such as laser power, laser resonance frequency, laser delay time and microwave power. 

The frequency and the splitting of the two resonance lines $f^{+}$ and $f^{-}$ are determined by scanning our laser system over a range of about 5~GHz centered around the two transition frequencies. The laser power was adjusted to observe a clear splitting of the two transition lines to ensure that only one of the two hyperfine levels of the $(n,L)=(36,34)$ state is depopulated by laser stimulation. 
The measurements were all performed at a target pressure of 250~mbar and a delay time of $T=350$~ns between the two lasers pulses. 
Due to limited measurement time, only one target density was used. However, previous studies in $\overline{\mathrm{p}}^{4}$He$^{+}$ \cite{Pask:2009} as well as calculations suggest that the target density should have no effect on the resonance line shape, width or amplitude of the resonance lines~\cite{Kman, Korenman:2009, Korenman:2012, Friedreich:2012} at the level of the precision of this experiment.

\begin{figure}[t]
\centering
\includegraphics[width=0.47\textwidth, trim=0 0 0 0,clip]{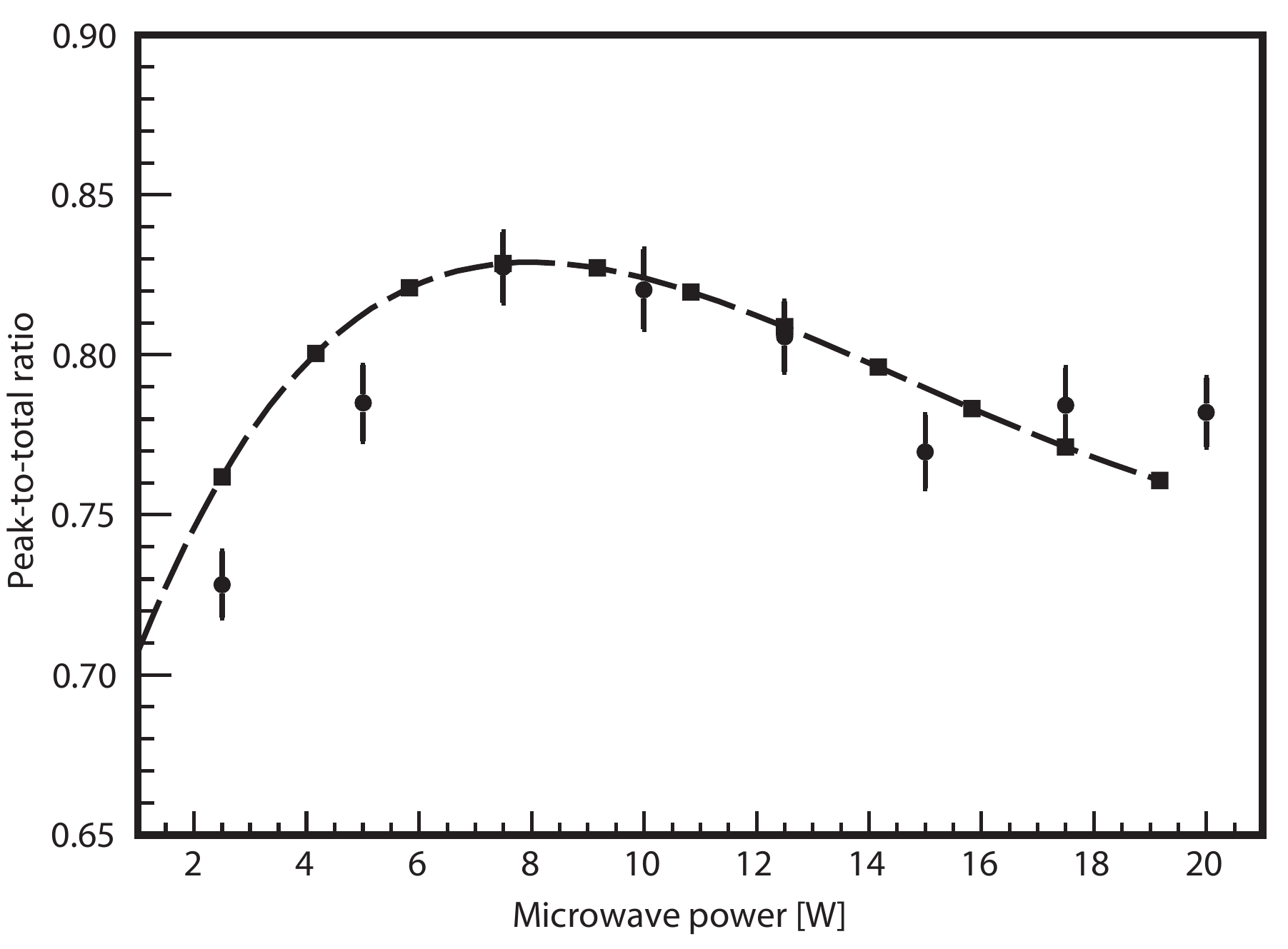}
\caption[Microwave Power Scan]{\small{Signal-to-noise ratio measured for several microwave powers  in comparison to a numerical simulation based on the used measurement parameters and normalised to the measured signal amplitude~\cite{Friedreich:2012}. The given power value is referring to the the power inside the target. 7.5~W were finally chosen for the microwave spectroscopic measurements.}} 
\label{fig:mwpowscan}
\end{figure}

It is important to choose the correct microwave power in order for the electron to undergo one electron spin-flip~\cite{friedreichNIM:2012}, i.e. to achieve a $\pi$-pulse that results in the highest signal. For this the ptt ratio is measured at the predicted resonance frequency for several power values in the range between 0 and 20 Watts microwave power inside the cavity, as determined by the pick-up antenna.
The points measured at 0~W were recorded on resonance. Points were also taken sufficiently off resonance (a few hundred Megahertz away) but at some non-zero power. Off resonance, the microwave pulse should have no effect on the atoms, thus confirming that the observed signal is real and not caused by some kind of fluctuations. 
Fig.~\ref{fig:mwpowscan} illustrates such a scan. According to these data, a $\pi$-pulse is completed at the first power maximum of about 7.5~W. The microwave power study is performed for a laser delay of 350~ns. 

\begin{figure}[h]
\centering
\includegraphics[width=0.49\textwidth, trim=0 0 0 0,clip]{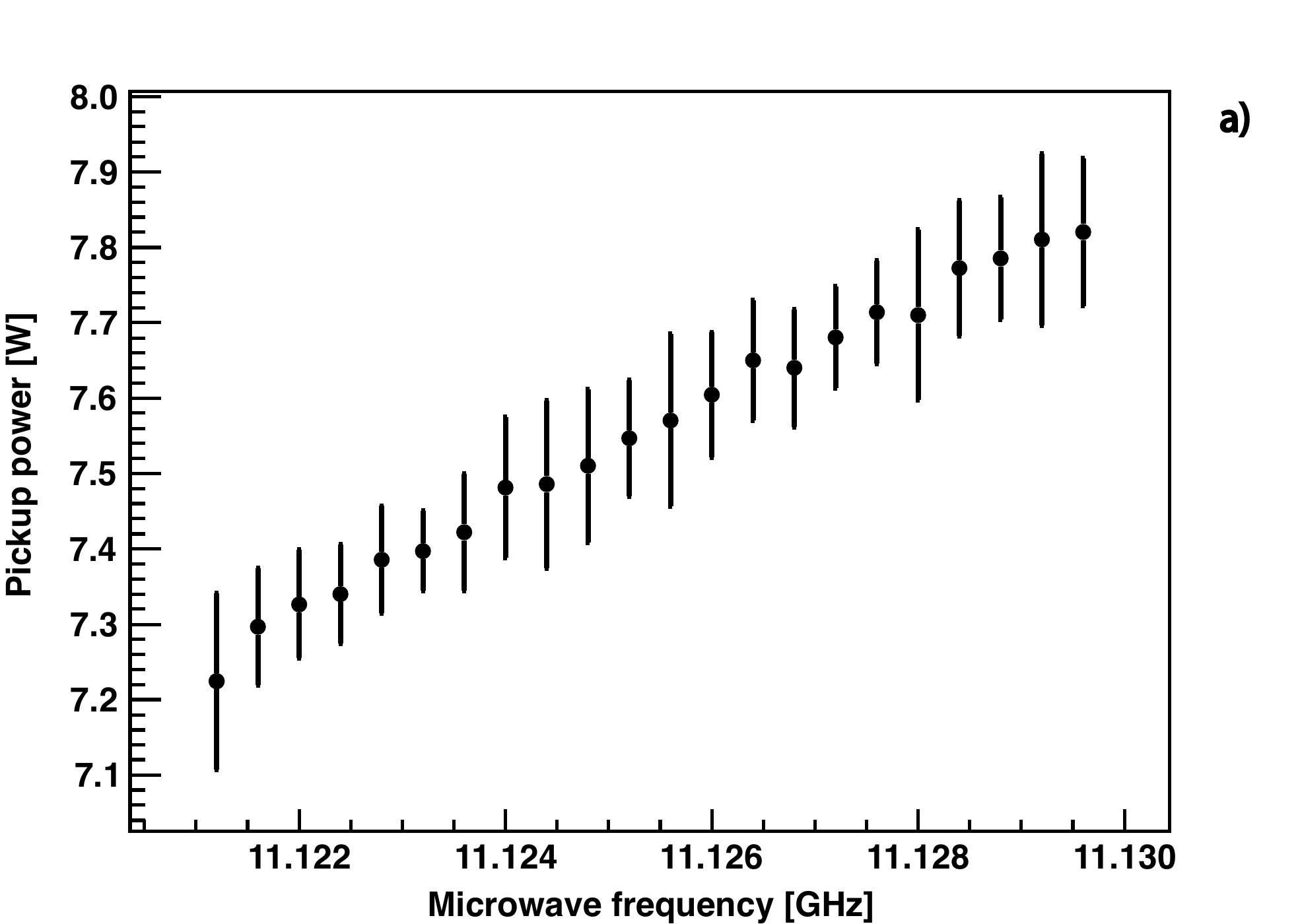}
\includegraphics[width=0.49\textwidth, trim=0 0 0 0,clip]{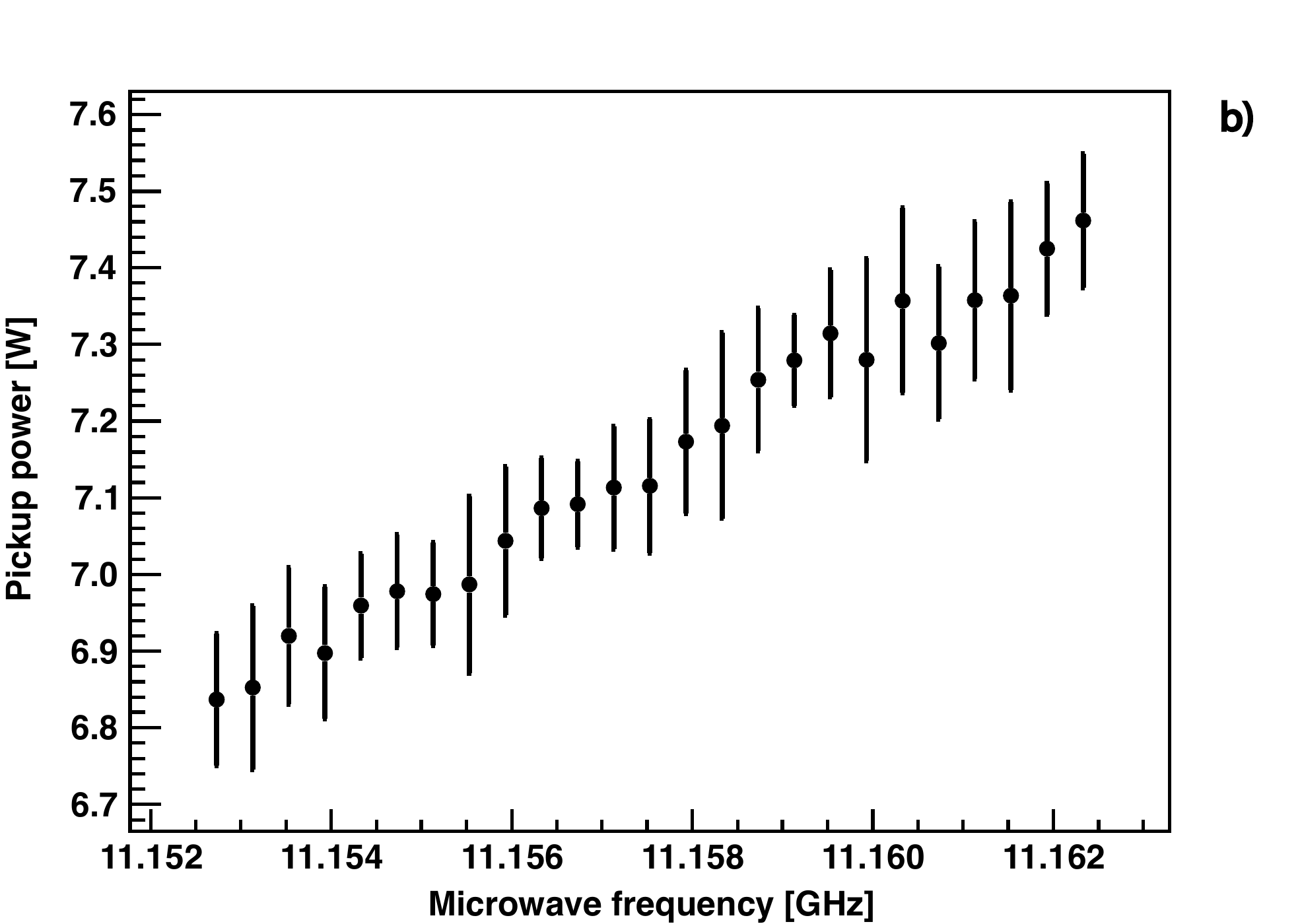}
\caption{\small{The change of the microwave power over the measured frequency range for the a) 11.125~GHz and the b) 11.157~GHz transition.}} 
\label{fig:mwpow}
\end{figure}

Figure~\ref{fig:mwpow} displays the frequency dependence of the microwave power over the scan range in the case of the two 11~GHz transitions -- with an average drift of 10-13\% over the recorded spectrum. Despite thorough calibration of the system, there appears to be a linear tendency of the power over the frequency range. This behavior could potentially lead to a distortion of the line shape and an increase of errors. However, from Fig.~\ref{fig:mwpowscan} can be seen that the peak-to-total ratio does not change considerably within the error between a microwave power of 7.5~W and 10~W. Therefore, it is not expected that this linear tendency of the power over the frequency range has a significant effect on the error and the fit of the transitions lines.

\subsection{The microwave transitions}
\label{mwtrans}
\begin{figure}[h]
\centering
\includegraphics[width=0.47\textwidth, trim=0 0 0 0, clip]{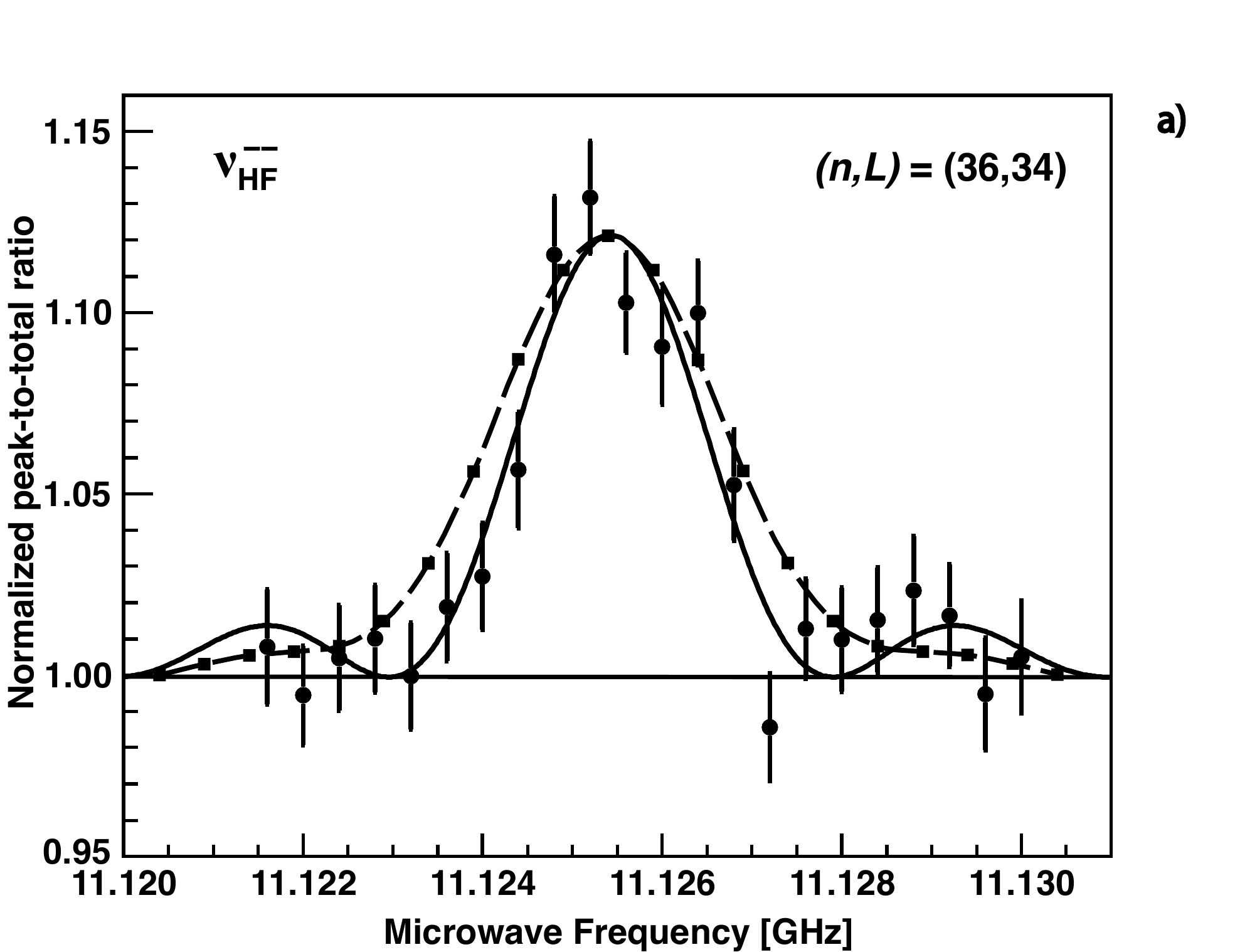} 
\includegraphics[width=0.47\textwidth, trim=0 0 0 0, clip]{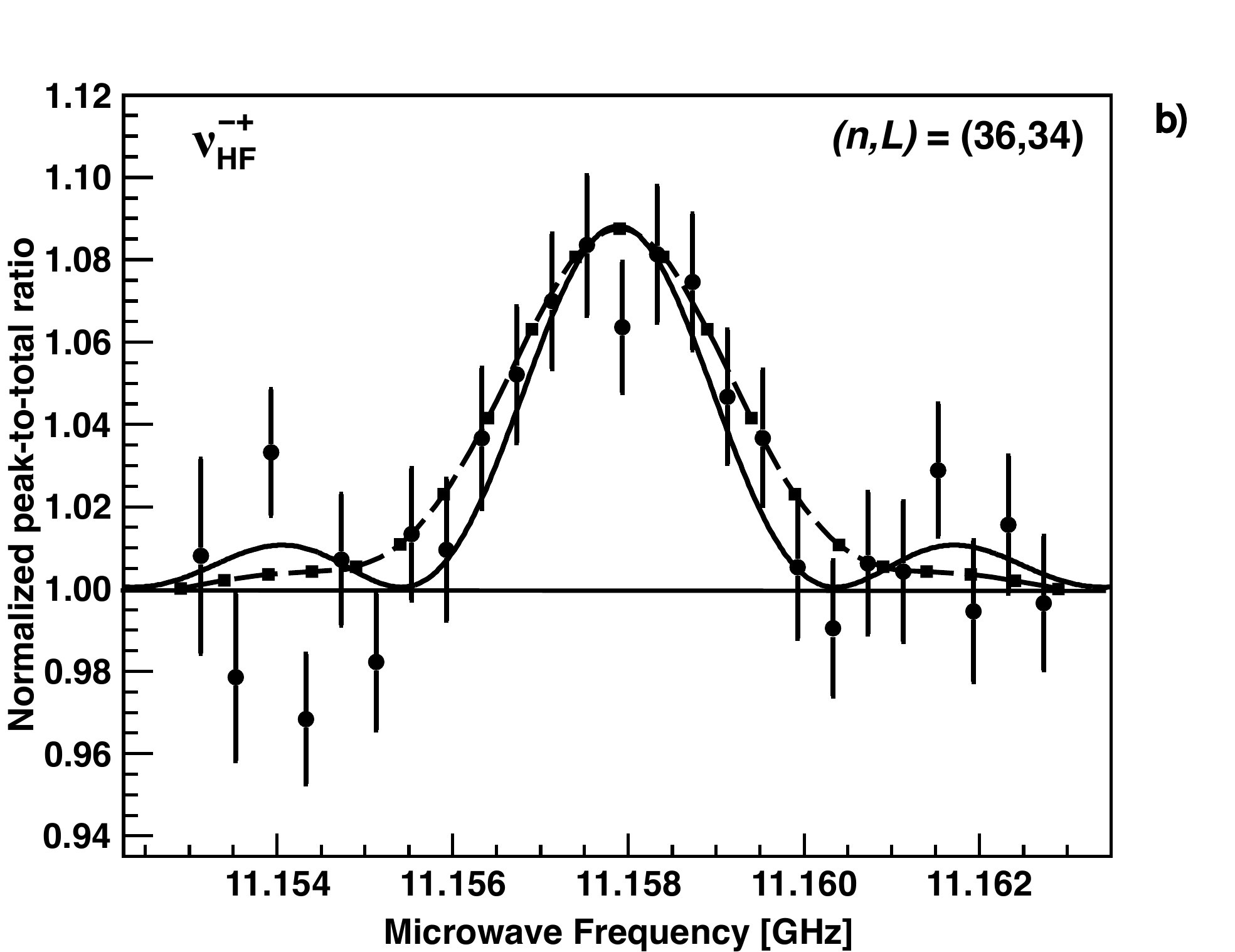} 
\caption{\small{Scan over the microwave frequency for the a) $\nu_{\mathrm{HF}}^{--}$ and the b) $\nu_{\mathrm{HF}}^{-+}$ transition of the $(n,L)=(36,34)$ state in $\overline{\mathrm{p}}^3$He$^+$, at a target pressure of 250~mbar, fitted with Eq.~\ref{eq:NatLineShape} (solid line) and using the simultaneous fitting of individual scans. The frequency of the measured transitions are $\nu_{\mathrm{HF}}^{--}=11.12548(08)$~GHz and $\nu_{\mathrm{HF}}^{-+}=11.15793(13)$~GHz. The dashed curve shows a simulation using collision rates obtained from comparison between experiment and simulation~\cite{Friedreich:2012}.}}
\label{fig:MWscan}
\end{figure}

Two of the four allowed {SSHF} resonance transitions in $\overline{\mathrm{p}}^3$He$^+$ could be observed. In the analysis all recorded data, including the previously published data of 2010 \cite{friedreich:2011} and new ones obtained in 2011 were taken into account. The two resonances were measured and fitted separately. For each microwave frequency scan 20-25 frequency points were recorded, equally spaced over a range of 9~MHz, centered around the theoretical transition frequency. Two analysis methods were used to average over data taken in different years and under different conditions: {\em average scan fitting (ASF)} in the case of identical conditions and frequency points, the data taken at the same frequency were first averaged using the method of weighted average, then the resulting scan was fitted. For {\em simultateous individual scan fitting (ISF)} the data points were not averaged but simultaneously fitted using the same values for central frequency and width but individual values for height or background levels. Using {ISF}, also scans taken with different microwave power or with different frequency points can be analyzed together. In the case of the $11.125$~GHz transition the value for every frequency was averaged over a total of 40 data points, for the $11.157$~GHz over a total of 42 data points.
These values were obtained using the simultaneous fitting of individual scans. The fit results are displayed in Fig.~\ref{fig:MWscan} in comparison with simulation curves.

\begin{center}
\begin{table}[b]
\caption{\small The table displays the uncorrected fit results $\nu_{\mathrm{HF}}^\mathrm{u}$ for the fitting of the raw data together with the reduced $\chi^2/ndf$ and $\nu_{\mathrm{HF}}$ after inflating the errors of the individual data points by $\sqrt{\chi^2/ndf}$. The fit transition frequencies are displayed for the two different fitting methods, {ASF} and {ISF}. At the higher resonance the frequency points differed slightly between  2010 and 2011. These data can only be combined in the averaging over all single scans. The  microwave power for the $11.157$~GHz resonance was further lower by about 2.5~W compared to 2011. 
Therefore, the values obtained by the ISF method were used as final results.}\label{tab:chisquare}

\hfill
\begin{tabular}[t!]{l@{}rrrr}
\br
Transition & Method & $\nu_{\mathrm{HF}}^\mathrm{u}$ (GHz) & $\chi^2/ndf$ & $\nu_{\mathrm{HF}}$ (GHz) \\ 
 \mr
$\nu_{\mathrm{HF}}^{--}$ & ASF & 11.12550(04) & 8.71 & 11.12550(08) \\  

$\nu_{\mathrm{HF}}^{--}$ & ISF & 11.12548(03) & 7.13 & 11.12548(08)  \\  

$\nu_{\mathrm{HF}}^{-+}$ (2010) & ASF &11.15830(07) & 8.26 & 11.15830(17)  \\  

$\nu_{\mathrm{HF}}^{-+}$ (2011) & ASF & 11.15760(07) & 8.42 & 11.15760(14) \\  
  
$\nu_{\mathrm{HF}}^{-+}$ & ISF & 11.15793(04) & 7.92 & 11.15793(13)\\  
\br
\end{tabular}
\hfill{}
\end{table}
\end{center}

To fit the two transitions, a function of the natural line shape for a two-level system which is affected by an oscillating magnetic field for a time $T$ was used. It is given by~\cite{Sflugge}
\begin{eqnarray}    
X(\omega) = A &\frac{\vert 2b \vert ^2} {\vert 2b \vert ^2 + (\omega_0 - \omega)^2} \nonumber \\
\times &\sin^{2} \left\lbrace \frac{1}{2} \left [ \vert 2b \vert ^2 + (\omega_0 - \omega)^2 \right ] ^{\frac{1}{2}} T \right\rbrace.
\label{eq:NatLineShape}         
\end{eqnarray}
Here $X(\omega)$ is the probability that an atom is transferred from one {HF} state to the other, $\omega$ is the angular frequency of the magnetic field and $\omega_0$ is the angular frequency of the transition between the two energy levels. $A$ is a scaling term which equals 1 in an ideal two-level system. In the fitting procedure this term takes into account that in reality the two-level system is not ideal. The parameter $b=\Omega/2$ is a time independent part of the transition matrix elements between two energy levels, with the Rabi frequency $\Omega(\mu B_0)/\hbar$ and $\mu$ denoting the calculated averaged magnetic dipole moment. The Rabi frequency is dependent on the microwave power. Using the calculated values for the average oscillating magnetic field amplitude of $B_0 = 0.24(4) \times 10^{-4}$~T  and the magnetic dipole moment, we obtain a Rabi frequency in the range of 10~MHz.
In the case of a complete $\pi$-pulse, one obtains $\vert b \vert T=\pi/2$. This is referred to as the optimum case, since together with $X(\omega)=1$ at resonance this gives the smallest width for the transition line, $\Gamma=0.799/T=2.28$~MHz for $T=350$~ns. 
The two observed microwave resonance transitions were measured and fitted individually with this function, adding a constant background. The side peaks in the fit are caused by the Rabi oscillations. From the fit, the frequencies for the measured $\nu_{\mathrm{HF}}^{--}$ and $\nu_{\mathrm{HF}}^{-+}$ transitions can be obtained. 

As seen in Tab.~\ref{tab:chisquare} the fit results of the scans show a normalized $\chi^2/ndf$ that is larger than one. This is a general feature of our analog method to measure the delayed annihilation time spectra: the ADATS consists of the digitized current output of the Cherenkov PMTs that does not directly carry the statistical information on the observed number of annihilations per time. From the observed fluctuations, the digitization error, and other parameters a error is calculated that systematically underestimates the fluctuations in the data. Therefore the errors of all data points of a scan are multiplied by $\sqrt{\chi^2/ndf}$ to obtain the correct errors of the fit results.

Regarding the errors, there are several systematic effects which had to be considered. The largest influence was due to shot-to-shot fluctuations of the antiproton beam. These effects were reduced by normalizing to the total intensity of the pulse and further normalizing the second annihilation peak to the first one. Therefore, mainly shot-to-shot fluctuations of the microwave power and deviations in the laser position and fluence from day to day -- although considerably smaller -- contributed to the error quoted in Table~\ref{tab:Results}. The individual contributions from fluctuations of antiproton beam and laser beam cannot be assessed separately. They are contained in the error obtained from the fit.

The laser power as well as the wavelength and the overlap between the two laser pulses were monitored and measured, concluding that the fluctuations give no relevant contribution to the measurement error. The mean laser energy changes by about 0.07\% over one measurement shift of eight hours, the laser wavelength drifts by about 0.002\%. It is difficult to quantify by how much fluctuations of the laser parameters influence the measured annihilation signal. 
\begin{center}
\begin{table}[b]
\caption{\small The experimental results for the $\nu_{\mathrm{HF}}^{--}$ and $\nu_{\mathrm{HF}}^{-+}$ in comparison with three-body {QED} calculations, where $\nu_{\rm{\mathrm{HF}}}$ denote the {SSHF} transition frequencies, $\delta_{\mathrm{exp}}$ is the relative error of the measured frequencies and $\Gamma$ the resonance line width. 
The relative deviation of experiment and theory is defined as $\delta_{\mathrm{th-exp}}=(\nu_{\mathrm{exp}}-\nu_{\mathrm{th}})/\nu_{\mathrm{exp}}$.
The quoted theoretical precision is $\sim 5 \times 10^{-5}$ from the limitation of the Breit-Pauli approximation that neglects terms of relative order $\alpha^{2}$. This does not include numerical errors from the different  variational methods used.
For ref. \cite{Kino:03APAC}  $\Delta \nu_{\mathrm{HF}}^{\pm}$  was calculated from the difference of the tabulated antiproton spin-flip transitions $J^{--+}\longrightarrow J^{---}$ and $J^{+-+}\longrightarrow J^{+--}$, resulting in an relative error of $3\times10^{-4}$.}
\label{tab:Results}
\begin{tabular}{l@{}rrrrrrrr}
\br
& $\nu_{\mathrm{exp}}$ & $\delta_{\mathrm{exp}}$ & $\Gamma$  & Korobov~\cite{Korobov:06,Korobov:2010} & $\delta_{\mathrm{th-exp}}$ & Kino~\cite{Kino:03APAC}& $\delta_{\mathrm{th-exp}}$ \\
& (GHz) & $\times10^{6}$ & (MHz) & (GHz) & $\times10^{6}$ & (GHz) & $\times10^{6}$ \\
\mr
$\nu_{\mathrm{HF}}^{--}$ & 11.125 48(08) & 7.2 & 1.69(11) & 11.125 00(56) & $43$ & 11.125 15(56) & 29 \\
$\nu_{\mathrm{HF}}^{-+}$& 11.157 93(13) & 11.7 & 2.20(15) & 11.157 73(56) & $18$ & 11.157 56(56) & 33 \\
\hline
&  & $\times10^{3}$ &  &  & $\times10^{3}$ &  & $\times10^{3}$ \\ \hline
$\Delta \nu_{\mathrm{HF}}^{\pm}$ & 0.032 45(15) & 4.7 &  & 0.032 721 9(16) & $-8.4$ & 0.032 408(11) &  $1.3$ \\
\br
\end{tabular}
\end{table}
\end{center}
The transition processes were numerically simulated by solving the optical Bloch equations in order to estimate important measurement parameters, in particular the required microwave power and the signal-to-noise ratio~\cite{Friedreich:2012}. The Bloch equations describe the depopulation of states, in this experiment induced by laser light and microwave radiation and under the influence of collisional effects. For most parameters, such as microwave power, {Q} value and laser delay, the measured values were taken. To assess the rates of collisional effects which induce relaxations between the {SSHF} states, the simulations are adjusted to the experimental results.
Two types of collisions can be distinguished - elastic and inelastic collisions. While elastic collisions can cause a broadening and shift of the resonance line, inelastic collisions will result in a spin exchange between the hyperfine substates which can lead to a decrease of the measured signal. Both, elastic collision rate $\gamma_e$ and inelastic collision rate $\gamma_i$, can have considerable systematic effects on the signal height, line shape and frequency of the transition line. 
The resulting calculated resonance curves are represented as dashed lines in Fig.~\ref{fig:MWscan}, showing good agreement with the experimental data. Extracting the elastic and inelastic collision rates $\gamma_e$ and $\gamma_i$ for the two transitions gives 
\begin{eqnarray}
\hspace{-2.1cm} \textnormal{for}~11.125~\textnormal{GHz}: \gamma_e^{--} = 3.45_{-0.71}^{+0.79}~\textnormal{MHz} \hspace{0.7cm} \textnormal{for}~11.157~\textnormal{GHz}: \gamma_e^{-+} = 3.48_{-0.99}^{+1.20}~\textnormal{MHz} \nonumber \\
\hspace{0.975cm} \gamma_i^{--} = 0.51_{-0.08}^{+0.09}~\textnormal{MHz} \hspace{3.8cm} \gamma_i^{-+} = 0.52_{-0.11}^{+0.13}~\textnormal{MHz}
\end{eqnarray} 
To obtain the errors for these rates, the annihilation signal amplitude was calculated for different values of the elastic and the inelastic collision rates for both transitions. The fitted annihilation signal amplitude of the transitions and its errors were then used to assess the collision rates for the minimum and maximum amplitude values within the $\pm 1 \sigma$ level by interpolation.

Based on theory, the collision rates are expected to be equal for different single electron spin flip transitions within a state~\cite{Korenman:2012}. To calculate the weighted mean of the values for the individual transitions, a mathematical model presented in~\cite{Barlow} which accounts for the asymmetric errors of the single values is used, resulting in an elastic collision rate of $\gamma_e = 3.41 \pm 0.62$~MHz and an inelastic collision rate of $\gamma_i = 0.51 \pm 0.07$~MHz. These rates go into the optical Bloch equations in the simulations as angular frequencies. Thus, in order to compare them to the total line widths $\Gamma=0.799/T$ of the measured resonances (see Table~\ref{tab:Results}), given as linear frequencies, they have to be divided by $2\pi$:
\begin{eqnarray}
\gamma_{e}^{'} = \frac{\gamma_e}{2\pi} = 0.54 \pm 0.10 ~\textnormal{MHz} \nonumber \\ 
\gamma_{i}^{'} = \frac{\gamma_i}{2\pi} = 0.08 \pm 0.01 ~\textnormal{MHz}.
\end{eqnarray}
Only the elastic collision rate affects the width of the resonance line while inelastic collisions affect the transition rate and thus the height of the resonance signal. 
The measured rates agree within a factor $2$ with theoretical calculations which obtain an elastic collision rate of approximately $0.48$~MHz and an inelastic collision rate of approximately $0.16$~MHz, given as linear frequencies~\cite{Korenman:2012}.

\begin{figure}
\centering
\includegraphics[width=0.8\textwidth, trim=40 0 0 0,clip]{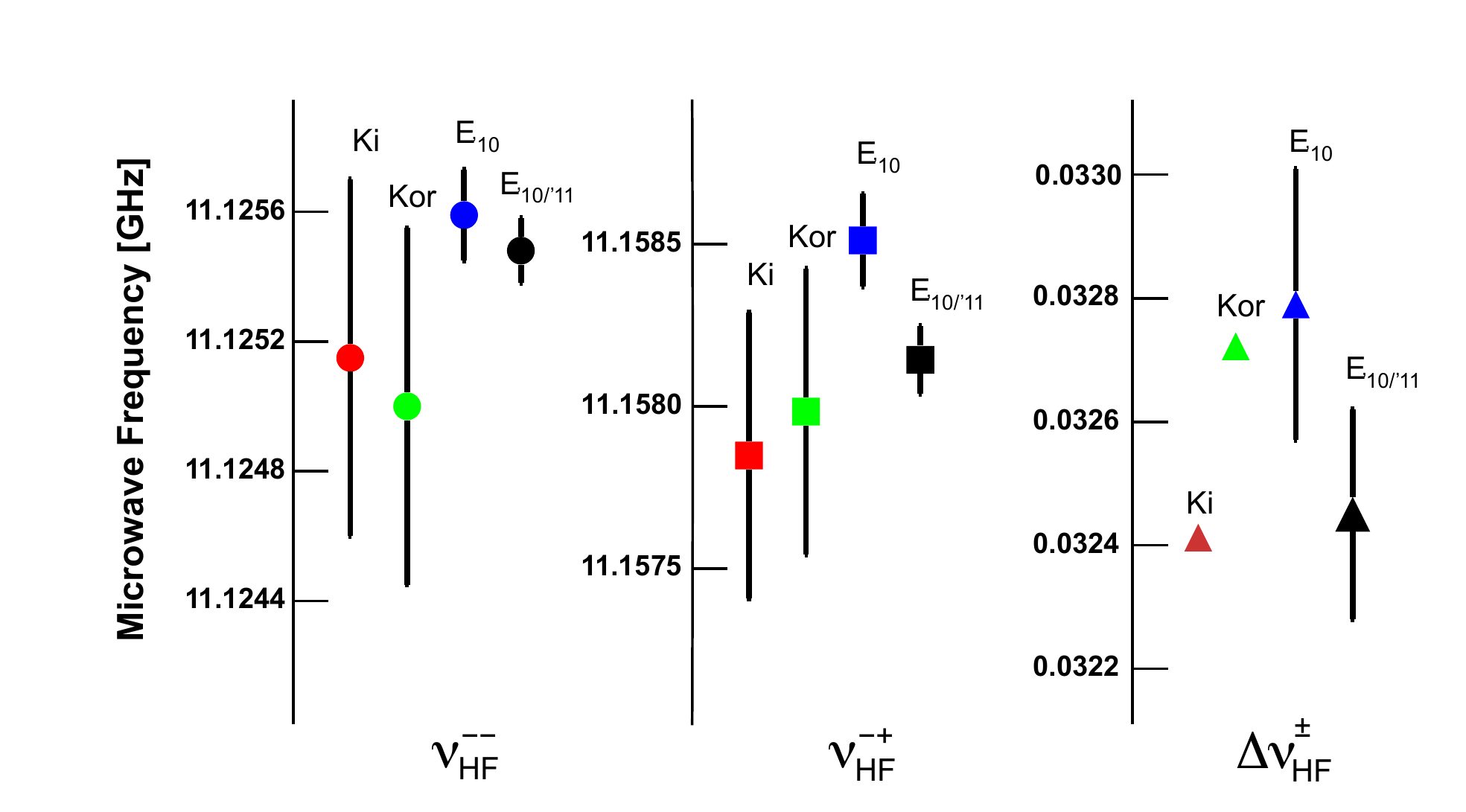} 
\caption[Summary of Principal Results]{\small{This graph summarizes the results for the two measured \textit{SSHF} transitions $\nu_{\mathrm{HF}}^{--}$ and $\nu_{\mathrm{HF}}^{-+}$ as well as the frequency difference $\Delta \nu_{\mathrm{HF}}^{\pm}$($\mathrm{E}_{'10}$~\cite{friedreich:2011}, $\mathrm{E}_{'10/'11}$) for the first measurement period in 2010 and the combined result of a all data recorded in the years 2010 and 2011. It further provides a comparison of these values with the respective theoretical calculations (Kor~\cite{Korobov:06, Korobov:2010}, Ki~\cite{Kino:03APAC}). The frequency difference of the experimental data for the $11.15773$~GHz resonance between the first year of measurements and the combined results of all recorded data may be explained by the slightly different microwave power used for the measurement period in 2010 and also by the lower statistics for this transition in the first year.}} 
\label{fig:MWoverview}
\end{figure}

\section{Conclusion}
\label{sec:concl}
Two of the four favored {SSHF} resonance transitions in $\overline{\mathrm{p}}^3$He$^+$ were observed and are in agreement with theory within the estimated theoretical error (cf. Tab.~\ref{tab:Results} and Fig.~\ref{fig:MWoverview}). The experimental errors have been decreased by 43\% for $\nu_{\mathrm{HF}}^{--}$ and 25\% for $\nu_{\mathrm{HF}}^{-+}$ compared to previously published results~\cite{friedreich:2011}. The value for $\nu_{\mathrm{HF}}^{-+}$ agrees better with theory than before. Also the frequency difference $\Delta \nu_{\mathrm{HF}}^{\pm}$ agrees with theoretical calculations. However, the experimental error for $\Delta \nu_{\mathrm{HF}}^{\pm}$ is still very large compared to theory. 

The measured hyperfine transition frequencies agree with theory within $0.2-0.5$ MHz (18--43 ppm). The current precision is still worse than for the most recent results with $\overline{\mathrm{p}}^4$He$^+$, which gave an error of 3~ppm for the individual transition lines~\cite{Pask:2009}. Due to limitations in antiproton beam quality this precision for $\overline{\mathrm{p}}^4$He$^+$ is not likely to be improved anymore. However, it is also unlikely to achieve an uncertainty for $\overline{\mathrm{p}}^3$He$^+$ transition frequencies as small as that for $\overline{\mathrm{p}}^4$He$^+$. There are eight instead of four {SSHF} energy levels in $\overline{\mathrm{p}}^3$He$^+$ and thus the measured signal will be only about half of the signal obtained for $\overline{\mathrm{p}}^4$He$^+$. Therefore much higher statistics would be required.

A comparison of the theoretical values for the two {SSHF} transitions at 11~GHz with the measurement results shows that there is a small shift in frequency towards higher values for both transitions (cf. Fig.~\ref{fig:MWoverview}). 
According to V. Korobov~\cite{Korobov:2010}, this discrepancy is most likely due to the theoretical limits of the Breit-Pauli approximation that has been used for the calculations. The relative error of the theoretical frequencies is estimated to be $\alpha^2=5 \times 10^{-5}$. 
The relative error of the theoretical frequencies is estimated to be $5 \times 10^{-5} \sim 0.56$~MHz. 
Together with the experimental error of $\sim$0.2~MHz there is agreement between experiment and theory. 

A density dependent shift could also contribute to this deviation. The density dependence is found to be much smaller for an {M1} transition, the electron spin-flip transitions induced by the microwave, than for an {E1} transition induced through laser stimulation~\cite{Pask:08}. In the case of $\overline{\mathrm{p}}^4$He$^+$ theoretical calculations of G. Korenman~\cite{Kman,Korenman:2009} confirmed that the density dependence is very small. Also for $\overline{\mathrm{p}}^3$He$^+$ theory predicts a collisional shift at the kHz level, much smaller than the experimental error bars~\cite{Korenman:2012}. 

For the frequency difference $\Delta \nu_{\mathrm{HF}}^{\pm} = \nu_{\mathrm{HF}}^{-+} - \nu_{\mathrm{HF}}^{--}$ between the two {SSHF} lines around 11~GHz there is an agreement between both theoretical results and experiment within 1.5 $\sigma$ of the experimental error of 150 kHz (0.47\%). $\Delta \nu_{\mathrm{HF}}^{\pm}$ is important due to its proportionality to the magnetic moment of the antiproton. The error of the theoretical value is 1.6~kHz, which is considerably smaller than the experimental error. The reason is that in theory the splitting between the transition lines can be calculated directly and the errors are the same for all transitions within the hyperfine structure whereas the experimental value of the splitting is received from the difference of the single transition lines. 

The two transitions at 16~GHz could not be measured anymore due to lack of beamtime -- even though the microwave target was readily tested and calibrated. However, we came to the conclusion that the observation of these two resonance lines would deliver no additional information on the investigated three-body system and primarily serve to accomplish a complete measurement of the $\overline{\mathrm{p}}^3$He$^+$ hyperfine structure.

This study with $\overline{\mathrm{p}}^3$He$^+$ was considered a test of {QED} calculations using a more complex system compared to $\overline{\mathrm{p}}^4$He$^+$ and thus provide a stronger confirmation of the theoretical models. With more statistics and careful investigation and accounting for systematic effects such as frequency dependencies of the single parts of the microwave setup the precision might realistically increase at most by a factor of two. Nonetheless, this would not reach the results achieved with $\overline{\mathrm{p}}^4$He$^+$ and thus not give a better experimental value for the antiproton magnetic moment, i.e. a better test of {CPT} invariance. Recently, the antiproton magnetic moment has been measured for the first time using a single trapped antiproton, reaching a precision of $4.4$~ppm~\cite{Gabrielse:2013} which is far outside the reach of the technique presented in this work.

With this study the spectroscopic measurements of the hyperfine structure of $\overline{\mathrm{p}}^3$He$^+$ are concluded. There are no further measurements planned. Based on the current experimental conditions no improvement of precision can be expected. Also the theory reached its limits using the calculation methods available at present. 

\ack

We are grateful to Dr. V. Korobov and Dr. G. Korenman for intensive discussions on the theoretical framework. Further, we want to thank our project students Matthias Fink, Johannes Handsteiner, Mario Krenn, Hans-Linus Pfau und Mariana Rihl for their help before and during the beamtime. This work has received funding from the Austrian Science Fund (FWF): [I--198--N20] as a joint FWF--RFBR (Russian Foundation for Basic Research) project, the Austrian Federal Ministry of Science and Research, the Japan Society for the Promotion of Science (JSPS), the Hungarian National Science Funds (OTKA K72172), the European Science Foundation (EURYI) and the Munich Advanced Photonics Cluster (MAP) of the Deutsche Forschungsgemeinschaft (DFG).

\end{linenumbers}
\end{document}